\documentclass[aps,preprint,noshowpacs,nofootinbib,superscriptaddress,groupedaddress, floatfix,superscriptaddress, preprintnumbers]{revtex4-1}  
 
\usepackage{graphicx} 
\usepackage{dcolumn}   
\usepackage{bm}        
\usepackage{amssymb}   
\usepackage{hyperref}

\usepackage{multirow}
\usepackage{feynmf}
\usepackage{amsmath,amssymb}

\usepackage{graphicx,bm}

\usepackage{slashed}

\usepackage{natbib}

\usepackage{epstopdf}

\usepackage{ulem} 

\usepackage[usenames]{color}

\usepackage{float}

\renewcommand\sout{\bgroup \color{red} \ULdepth=-.5ex \ULset}

\newcommand{\be}{\begin{equation}}
\newcommand{\ee}{\end{equation}}
\newcommand{\ba}{\begin{eqnarray}}
\newcommand{\ea}{\end{eqnarray}}

\usepackage{soul} 

\newcommand{\bea}{\begin{eqnarray}}
\newcommand{\eea}{\end{eqnarray}}

\def\d{\delta}
     \def\t{\tau} \def\o{\omega}\def\r{\rho}
\newcommand{\eq}[1]{Eq.~(\ref{#1})}\newcommand{\half}{{\frac{1}{2}}}
\def\xtr{x_{\rm trans}}

\begin{document}
\begin{flushright}
\end{flushright}

\preprint{NT@UW-24-07}

\title{The Quark Pauli Principle and the Transmutation of Nuclear  Matter} 


\author{Larry McLerran }
\affiliation{Institute for Nuclear Theory, University of Washington, Box 351550, Seattle, WA 98195, USA}
\email{mclerran@me.com}

\author{Gerald A. Miller}
\affiliation{Department of Physics, University of Washington, Seattle, WA 98195, USA}
\email{miller@uw.edu}

\date{\today}

\begin{abstract}
The phase space density, $\r^Q$, of quarks in nuclei is studied using realistic models of unintegrated quark distributions, known as transverse momentum densities (TMDs). If this density  exceeds unity for matter at normal nuclear densities,  the effects of the quark Pauli principle must play a role in nuclei, and models in which the nucleon density at low momentum is  small (Quarkyonic matter) may become a starting point for an entirely new description of nuclei. We denote the nuclear density for which $\r^Q=1$ to be a transmutation density, $n_T$, because quark degrees of freedom must be relevant at that density. Including the   TMDs of [G. de Teramond et. al, \href{DOI:https://doi.org/10.1103/PhysRevLett.120.182001} Phys. Rev. Lett. {\bf 120}, 182002, (2018)] for the valence quarks and phenomenological TMDs for the sea quarks we find that $n_T=0.17 \pm 0.04\,\rm fm^{-3}$, the density of normal nuclear matter.
Some of fhe implications of this finding are discussed.

\end{abstract}

\pacs{}

\maketitle

{\bf Introduction}

In Ref.\cite{McLerran:2007qj}, a class of solvable models of Quarkyonic matter, corresponding to a filled quark Fermi sea with a baryonic Fermi surface were proposed.  These Ideal Dual Quarkyonic (IdylliQ)  models correspond to a free gas of nucleons composed of quarks with a constraint that the phase space density of quarks cannot exceed one~\cite{Fujimoto:2023mzy}.  The phase space density of quarks was determined by the probability to find a quark inside a nucleon convoluted with the nucleon density.  For a particular choice of probability density, this model was explicitly solved.  The IdylliQ models are constructed to be explicitly dual between the nucleon and the quark descriptions.  There are two phases.  The low density phase is an ideal gas of nucleons and nowhere does the quark phase space density equal or exceed one.  On the other hand, the high density phase corresponds to a filled Fermi sphere of quarks with a diffuse Fermi surface.  The nucleon sits on a shell in momentum space at a momentum of order $k_F^N \sim N_c k_F^Q$ where $k_F^Q$ is the Fermi momentum of the filled quark Fermi sea and $N_c$ is the number of quark colors.  In addition to the shell of nucleons at the Fermi surface, there is an under-occupied density of nucleons which corresponds to the Fermi sea of quarks. The onset density where this occurs is determined from the probability density to find a quark inside a hadron at zero quark momentum becoming equal to one. In Quarkyonic matter, a hard equation of state, $P \sim {1\over 3} \epsilon$, is rapidly approached~\cite{McLerran:2018hbz}.

In Ref. \cite{Koch:2024zag}, the Quarkyonic model of noninteracting nucleons was generalized to include pion and sigma meson interactions.  A choice of the probability distribution to find a quark inside the nucleon was made that allowed an exact solution of the theory, with scalar mesons  ($\sigma$) and $\pi$ exchange interactions included in the mean field approximation. 
When the momentum scale in this probability distribution, and the  scalar meson coupling constants were tuned to produce the binding energy and density of isospin symmetric nuclear matter, it was found that the Quarkyonic transition occurred slightly below the density of nuclear matter, and an acceptable phenomenology of the bulk properties of nuclear matter was extracted.

A simple relationship that determines the onset density of Quarkyonic matter was found in Ref.  \cite{McLerran:2007qj}   and was exploited in Ref.~\cite{Koch:2024zag}.  This relation is best expressed in terms of phase space densities.  The phase space density for a particle is determined from its number density as
\begin{equation}
  \rho(k) = {1 \over N_{dof}}  (2\pi)^3 {{dn} \over {d^3k}}
\end{equation}
The factor of $N_{dof}$ accounts for the number of states due to degeneracy in color, spin and isospin. We  assume degeneracy in these degrees of freedom and will consider isosinglet nuclear matter. 
The phase space density corresponds to the quantum occupation number of states at a momentum $k$, and with the factor of $N_{dof}$ extracted, will have a maximum value of 1 for fermions.
  The phase space density of quarks at finite temperature and density system of quarks would be $\rho^Q(E) = {1 \over {1+e^{(E-\mu_Q)/T}}}$ where $T$ is the temperature and $\mu_Q = \mu_N/N_c$ is the chemical potential for the quarks, and there is a similar relation for nucleons, 
  $\rho^N(E) = {1 \over {1+e^{(E-\mu_N)/T}}}$. 
     If we ignore the anti-quarks, the baryon number density $n_B$ satisfies,
\begin{equation}
  n_B = 4 \int~ {{d^3p} \over {(2\pi)^3}} ~\rho^Q(p) =  4\int~ {{d^3p} \over {(2\pi)^3}}~ \rho^N(p)
   \end{equation}
because quarks carry baryon number $1/N_c$ and the contribution from the quark phase space density has an overall factor of $N_c$.  When anti-quarks are included in the nucleon wave function, the above sum rule represents  the difference between quark and antiquark phase space densities. 
Including anti-quarks does not affect the value of $\rho^Q -\rho^{\overline{Q}}$, but when we relate phase space densities of nucleons to quarks, $q \overline{q}$ pairs will play a role in determining the phase space density of quarks, $\rho^Q$.  The phase space density represents the occupation number of quantum mechanical states, and the Pauli principle imposes a restriction $\rho^Q \le 1$.  The present work is concerned with the zero temperature limit.

Ref.~\cite{Koch:2024zag} assumed there are only valence quarks in the nucleon. In that work,  the quark and nucleon phase space densities are related as
\begin{equation}
 \rho^Q(\vec k) = \int {{d^3p} \over (2\pi)^3} K(\vec k-\vec p/N_c) \rho^N(p)
\end{equation}
where $K$ is the normalized probability distribution to find a quark inside a nucleon,
$  \int~ {{d^3k} \over {(2\pi)^3}} K(\vec k) = 1$. 
This quark distribution saturates when $\rho^Q(0) = 1$, and an equation that determines the onset density was obtained.  The content of Ref.\cite{McLerran:2007qj} was to analyze this IdylliQ model  for a system of free nucleons composed of quarks satisfying the constraint that $\rho^Q \le 1$.  There is a transition from an ideal gas of nucleons to Quarkyonic matter when this condition is first satisfied, $\rho^Q(k = 0) = 1$.  In Ref. \cite{Koch:2024zag}, it was argued that this transition might occur at densities below nuclear matter and then assuming this is the case, with the inclusion of interactions with a pion and  sigma field, a reasonable phenomenology of nuclear matter was constructed. An unusual nucleon occupation density in which $\rho^N$ is $1/N_c^3$ for small nucleon momenta was obtained. A viable description of quasi-elastic electron scattering from nuclear matter was obtained despite this exceptional feature. 

The purpose of this paper is to derive a condition for the onset of Quarkyonic matter using experimentally constrained probability dsitributions for quarks and anti-quarks in a nucleon.  In particular, we will use the Light Front QCD distribution functions of the HLFHS collaboration\cite{deTeramond:2018ecg},\cite{Sufian:2016hwn}, as this formalism provide a 
very
 successful hadron phenomenology  with specific  forms of  distribution functions. 
We also will employ the sea quark distributions of CTEQ10~\cite{Lai:2010vv}.

{\bf Quark phase space density}



We use light-front, Fock-Space wave functions to obtain $\r^Q(\vec k)$~\cite{Brodsky:1997de}. In such a description this quantity 
is the square of a light-front wave function. Such a wave function depends on a longitudinal momentum fraction $x$ and the transverse component of momentum, $\vec k_T$, {\it e.g.} $|\psi(x,\vec k_T)|^2$.  These quantities are commonly called unintegrated distributions or transverse momentum distributions~\cite{Angeles-Martinez:2015sea}.

One may relate $\r^q(\vec k)$ to $\r^q(x,\vec k_T)$ by using the transformation
\be 
x= {k_z+E(\vec k)\over p^+}
\label{xdef}
\ee
where the $p^+$ is the sum of the nucleon's energy and $z$-component of momentum, and ${\partial x\over \partial k_z}=x/p^+.$ Preserving the integral over all values of $\vec k$ leads to 
the equivalent expressions:
${d^3n_q\over d^3k}= {x \over E} {d^3n_q\over dxd^2k_\perp}$ so that
\bea
 n_q(\vec k)={x\over E  } {d^3n\over dx d^2k_\perp}={x\over E } |\psi(x, \vec k_T)|^2 
 \label{nqa}
\eea
We shall be guided by Ref.~\cite{deTeramond:2018ecg} that presents models of wave functions for specific Fock-space components of valence quark  wave functions. These do not carry the entire momentum of the nucleon. It is well-known that gluons and the sea carry the remainder.  Therefore, we  use that paper's valence quark distributions and include also the effects of the quark sea using the CTEQ10 distribution functions.


{\bf Light-front wave functions}

The light-front valence wave functions of~\cite{deTeramond:2018ecg} are given for  parton numbers $\t=3,4$   by
\be  \label{LFWFk}
\psi_{\rm eff}^\tau(x, \vec{k}_T) 
= 4\sqrt{2} \pi \frac{ \sqrt{q_\tau(x) f(x) }}{1-x} \,
\exp\left[ -  \frac{2 f(x)}{(1-x)^2} \, \vec{k}_T^2 \right],
\ee
with normalization $\int_0^1 dx \int \frac{d^2\vec{k}_T}{(2 \pi)^3} \left\vert  \psi_{\it eff}(x, \vec{k}_T) \right\vert^2=1.$ 
The quark distribution functions are given by
\bea \label{qx} 
& q_\tau(x) =\frac{\Gamma(\t-1/2)}{\Gamma(\t-1)\sqrt{\pi}} \big(1- w(x)\big)^{\tau-2}\, w(x)^{- \half}\, w'(x)\\
& w(x) = x^{1-x} e^{-a (1-x)^2},
\eea
with. $f(x)={1\over 4\lambda} \log{1/w(x)}$, and
 $a$ a flavor independent parameter $a = 0.531 \pm 0.037$,  and  $\lambda =(0.548\,{\rm GeV})^2$.
 The valence $u$ and $d$ quark distributions are given by:
\bea
u_{\rm v}(x) &=&{3\over 2} q_{\tau=3}(x) +\frac{1}{2} \, q_{\tau=4}(x), \label{ux}\\
d_{\rm v}(x) &=&  \, q_{\tau=4}(x) \label{dx} ,
\eea
The normalization is 
$ \int_0^1 dx \, q_\t (x) = 1$. 
These distributions are in good agreement with experimental values and the nucleon electromagnetic form factors derived from these wave functions are also in good accord with measured data.  

These considerations enable us to develop the necessary more detailed versions of \eq{nqa}, to be able to use  parton distribution functions and TMDs that are functions of light cone variables.   We rewrite \eq{nqa} using $E(\vec k)=\sqrt{\vec k^2+ \o_q^2}$, where $\o_q$ is a suitable quark energy. In the non-relativistic quark model $\o_q$ is a quark mass and in the bag model it is a mode energy. 

Then with $\vec k_T=0$ and \eq{xdef}, one finds
\bea k_z={x^2 M^2 -\o_q^2\over 2 x M},
\label{kofx}\eea
so that $k_z(\t)=0$ corresponds to  a transmutation value of $x_{\rm trans}=\o_q/M$. 
The function $\o_q$ depends on the given Fock-Space component, so that  $x_{\rm trans}$ varies with the component even though 
$\vec k$ vanishes
for all components. 

This result together with Eqn. \ref{nqa}, gives  the quark  probability density per unit momentum at $k = 0$ where $x_0$ refers to the average $x$ value for a particular Fock space component as
\be
  n_q(\vec k = 0) = {1 \over M} |\psi(x_{\rm trans}, \vec k_T)|^2 \label{nq}
\ee
It is interesting to note that this result would be obtained immediately by considering the phase space density to be ${d^3n\over dk^+ d^2k_\perp} $ with $k^+=x M$.

The next step is to estimate the values of $x_{\rm trans}$. We assume that the glue accounts for one-half of the nucleon mass. Then for $\t=3$ and 
 taking all Fermions to have an equal contribution to the nucleon mass, we find $\o_q(\t=3)={1 \over 6} M,$ so that $\xtr(3)= {1 \over 6} $ for that component.
 The   component with 
$\t=4$   is a valence contribution in the theory of Ref.~\cite{deTeramond:2018ecg}, so that if we consider the system to be three quarks along with a valence gluon, we find $ \o_q(\t=4)={1 \over 8} M,$ so that $\xtr(4)= {1 \over 8}$.
It is useful to compare these values with the average values of $x$ computed using $q_{3,4}$ of \eq{qx}. Defining $x_\t\equiv \int_0^1 dx\, q_\t(x) $ we find $x_3=0.184$ and $x_4=0.136$ in good agreement with the estimates $1/6$ and
$1/8$. This means that on  average  the values of $x$ are close to $\xtr$. Using \eq{ux} and \eq{dx} we find the average value of  $x$ for  the up quark is 0.172 and that for the down quark is 0.126. These are in good accord with the values from CTEQ10 of 0.153 and 0.126. 

The formalism of~\cite{deTeramond:2018ecg} is not designed to give expressions for sea quark distributions. Instead, we modify  \eq{LFWFk} by replacing  $q_\t$ of   \eq{qx} by sea quark distributions $u_s(x)$ and $d_s$ taken from CTEQ10. This procedure follows  the logic of ~\cite{deTeramond:2018ecg}   because the  dependence on $\vec k_T$ is universal in that logic.
Doing this maintains the spatial extent of the configuration to of the nucleon size. 
We next need  to apply \eq{kofx}, again assuming that sea gluons account for half of the nucleon mass.  Three quarks plus one $q\bar q$ pair would lead to a five-parton system and $x=1/10.$ This value is an upper limit, since this is the minimal state which has quark-antiquark pairs in it.  Other possible  values of examined below. 

{\bf Nuclear densities}

 The next step is to place the nucleon into the nucleus.
The standard convolution formula, when applied to  the quark phase space density  reads
\bea n_q^\tau(x,\vec k_T)={1\over M } \int_x^\infty{ dy\over y}  |\psi_{\rm eff}^\tau(x/y, \vec{k}_T)|^2f_N(y),
\label{nq}\eea  
where $\psi_{\rm eff}^\tau(x/y, \vec {k}_T) $ is the light-front wave function for the given Fock-space component and $f_N(y)$ is the probability that a nucleon carries a momentum fraction $y= (E(p)+p_z/ M$, where $\vec p$ is the nucleon momentum and $M$ is the nucleon mass.
The derivation of this formula for the integrated (over $\vec k_T$) is discussed  in many reviews~\cite{Geesaman95,Norton:2003cb,Hen:2013oha,Hen:2016kwk}. One simple derivation, based on Feynman diagrams  is in ~\cite{Miller:2001tg}.

    The function $f_N(y)$ is given by 
    \bea
    f_N(y)=\int {d^2p_\perp dp^+ \over  (2\pi)^3}\d(y-{p^+\over M}) n_N(\vec p)
    \eea 
      with $p_z={1\over 2} (p^+ - {{ M}^2+p_\perp^2\over p^+})$.  In the present work we take the nucleon momentum density  $n(p)= \Theta(k_F^2-p_\perp^2-p_z^2)$, a standard Fermi gas function. In that case $\int dy f_N(y)=1/(6 \pi^2)k_F^3$, where $k_F$ is the Fermi momentum and the integral is one-fourth of the nucleon density of nuclear matter, $\rho_{NM}/4$. For such a density the integral above is approximated by using $f_N(y)=\delta(1-y)$, accurate to better than  1 \% for the values of $x$ that are of interest here.  Thus  for a given value of $\o_q$:
      \bea  \r_q(\vec k=0)={1\over M } |\psi_{\rm eff}^\tau({\o_q\over M} ,\vec{k}_T=0)|^2 { \rho_{NM}\over 4}
      \label{nq}\eea

{\bf Pauli principle condition}

We are concerned with isospin-symmetric nuclear matter ($N=Z)$, so that the isoscalar wave function  at $\vec k=0$ is given by
 \bea |\psi_v|^2\equiv  {1 \over 2}(|\psi_{\rm eff}^3({1\over 6},0)|^2+|\psi_{\rm eff}^4({1\over 8},0)|^2), \label{v}\eea 
 where the arguments are obtained from \eq{kofx}.
 The isoscalar wave function for the sea is given, in a first estimate, by 
 \bea |\psi_s|^2(x)\equiv {1 \over 3}(|\psi_{\rm u\, sea}({x},0)|^2+|\psi_{\rm d\, sea}({x},0)|^2),  \eea
  with the factor 1/3 accounting for the fact that there are three colors of quarks.   As noted above,  the sea could contain many anti-quarks. A 7 fermion system would have an average value of $x$ of 1/14=0.07, and a 9 quark system would have an average value of $x$ of 1/18=0.055. We  taking the  probability of having between 5 and 7 fermions to be evenly distributed so that, in  a first estimate,  we use 
\bea
\overline{|\psi_{\rm s\, }|^2}\equiv{1\over3} ( |\psi_{\rm s}({.1},0)|^2+ |\psi_{\rm  s}({1/14},0)|^2
+ |\psi_{\rm s}({1/18},0)|^2).\label{ave}\eea

\eq{nq} shows that the quark number density increases as the value of $\r_{NM}$ increases. Since $\r^q\le 1$ there will be a maximum value of $\r_{NM} $ that causes $\r^q(\vec k=0)$ to be unity. Let us define that critical value of nuclear density, for which the quark Pauli principle  is unity to be the transmutation density  $\r_T$. 
Then \eq{nq} tells us that
\bea \r_T^{\rm in}= \frac{4M}{ |\psi_v|^2+\overline{ |\psi_s|^2}},\label{rT}\eea
where the superscript refers to an initial estimate
The standard value of the density of nuclear matter is $\r_0=0.17 \,\rm fm^{-3}$ ~\cite{Bohr1969}. If $\r_T\ge \r_0$ then the quark Pauli principle must enter into the physics of nuclei.

Using \eq{rT} leads to the result that  $\r_T^{\rm in}=0.168=\r_0$, so that normal nuclear matter may be quarkyonic!  In this calculation the valence quark contribution to the denominator of \eq{rT} is about 55\% of the total and the sea quark contribution is 45\%.
The valence quark contribution to $\r_T$ is consistent with the estimate obtained in Ref.~\cite{Koch:2024zag} using a Gaussian model of the nucleon momentum density constrained by the measured charge radius of the proton.

It is necessary to account for the uncertainty that arises  from  using different estimates of $\o_q$.
A  rough estimate of the uncertainty in this estimate is obtained by using 
the spread of $x-$values between $1/10$ and $1/18$ instead of the average in \eq{ave}. This gives
\be
\r_T\approx 0.17 \pm 0.04\,\rm fm^{-3}.\ee
We also examine the variation caused by varying the value of 1/6 and 1/8   appearing by by plus or minus 10 \% 
to account for differences between the theory and CTEQ10 for the valence quarks. The effect of this variation along with that of varying the value of $x_s$ at which the sea quark distributions is  shown in Fig.~1. The central result is that it is reasonable to expect that the quark Pauli principle plays a role in nuclear physics at normal densities. Normal nuclear matter might very well be quarkyonic.

\begin{figure}[h]
  \centering
 \includegraphics[width=0.4\textwidth]{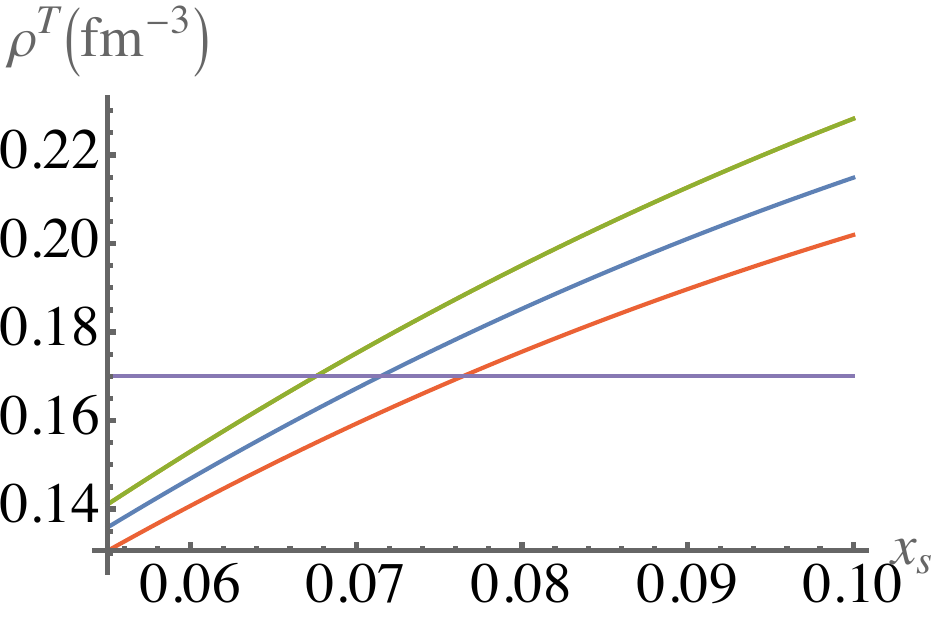}
  \hspace{0.3cm}
  \caption{ Variation of transition density, $\r^T$ with  $x_s$. The three different curves are obtained from \eq{rT} and then varying the arguments of the valence quark distributions by $\pm 10 $\%.
    }
\label{Mmod}  \end{figure}

{\bf Implications and Discussion}

In this paper, we determined the transmutation density to Quarkyonic matter using phenomenologically justified transverse momentum distributions,  parton distribution functions  unintegrated in $k_T$.  This advances  the treatment of Ref.  \cite{Koch:2024zag}, which used a probability distribution, $K(k,p)$,  chosen to simplify the computations, might  be criticized for lack of support from direct measurements of such distributions.  Our computations here strongly suggest that the transmutation density is quite close to that of nuclear matter at normal density, even though there are significant uncertainties in evaluating the kinematics at which $\vec k=0$  holds. 

Suppose  the Pauli repulsion of quarks, that enters when $\r^T$=1, is indeed  the origin of the mechanism which stabilizes nuclear matter against collapse. This suggests that $\o$ meson repulsion or repulsive contact interactions,  the stabilization mechanisms of many nuclear computations, has its origin in the Fermi repulsion of quarks arising from the Pauli principle.  The treatment of such repulsive forces  results  \cite{Koch:2024zag} in much different Fermi distributions of nucleons than is the case
for mean field treatments of nuclear matter.  This is because the overlap of quarks in the core leads to a filling of quark energy levels, which imply that the corresponding nucleon distributions will be under-occupied.  This leads to a shell-like structure for the momentum distributions of nucleons.  The resulting physical picture of nuclear matter has fundamental differences from that of a mean field description. Indeed, nuclear matter might not be a normal Fermi liquid.

We recognize that the ideas of this paper must be tested against existing measurements of nuclear properties. 
 Ref.  \cite{Koch:2024zag} already  shows that Quarkyonic matter is not inconsistent with data taken in the region of the quasi-elastic peak of electron scattering, but further exploration is necessary.  Quantum fluctuations of nuclear density must be considered. Short-ranged correlations occur when two nucleons overlap in space. If so, the density can be  higher than $\r_0$, and the effects of quark degrees of freedom must be relevant {\it i.e.} short-ranged nucleon-nucleon correlations must be associated with quark degrees of freedom.  The influences of quarks  cause the nuclear modification of parton distribution functions, the EMC effect~\cite{Geesaman95,Norton:2003cb,Hen:2013oha,Hen:2016kwk}.  A much more detailed analysis 
 of the relationship between Quarkyonic matter and the EMC effect is warranted. Nevertheless, it is fair to infer that the  considerations of this paper  strengthen the idea that there is a causal relation between  between the EMC effect and short-ranged correlations as in  the review~\cite{Hen:2016kwk}.
 
\section*{Acknowledgements}
L.M. acknowledges the support by the US Department of Energy under contract number DE-FG02-00ER41132.
  and G.A.M. acknowledges the DOE grant No. DE-FG02-97ER41014. We thank Dmitriy N. Kim  for providing the CTEQ10 quark distributions.


\begin{thebibliography}{15}%
\makeatletter
\providecommand \@ifxundefined [1]{%
 \@ifx{#1\undefined}
}%
\providecommand \@ifnum [1]{%
 \ifnum #1\expandafter \@firstoftwo
 \else \expandafter \@secondoftwo
 \fi
}%
\providecommand \@ifx [1]{%
 \ifx #1\expandafter \@firstoftwo
 \else \expandafter \@secondoftwo
 \fi
}%
\providecommand \natexlab [1]{#1}%
\providecommand \enquote  [1]{``#1''}%
\providecommand \bibnamefont  [1]{#1}%
\providecommand \bibfnamefont [1]{#1}%
\providecommand \citenamefont [1]{#1}%
\providecommand \href@noop [0]{\@secondoftwo}%
\providecommand \href [0]{\begingroup \@sanitize@url \@href}%
\providecommand \@href[1]{\@@startlink{#1}\@@href}%
\providecommand \@@href[1]{\endgroup#1\@@endlink}%
\providecommand \@sanitize@url [0]{\catcode `\\12\catcode `\$12\catcode
  `\&12\catcode `\#12\catcode `\^12\catcode `\_12\catcode `\%12\relax}%
\providecommand \@@startlink[1]{}%
\providecommand \@@endlink[0]{}%
\providecommand \url  [0]{\begingroup\@sanitize@url \@url }%
\providecommand \@url [1]{\endgroup\@href {#1}{\urlprefix }}%
\providecommand \urlprefix  [0]{URL }%
\providecommand \Eprint [0]{\href }%
\providecommand \doibase [0]{http://dx.doi.org/}%
\providecommand \selectlanguage [0]{\@gobble}%
\providecommand \bibinfo  [0]{\@secondoftwo}%
\providecommand \bibfield  [0]{\@secondoftwo}%
\providecommand \translation [1]{[#1]}%
\providecommand \BibitemOpen [0]{}%
\providecommand \bibitemStop [0]{}%
\providecommand \bibitemNoStop [0]{.\EOS\space}%
\providecommand \EOS [0]{\spacefactor3000\relax}%
\providecommand \BibitemShut  [1]{\csname bibitem#1\endcsname}%
\let\auto@bib@innerbib\@empty
\bibitem [{\citenamefont {McLerran}\ and\ \citenamefont
  {Pisarski}(2007)}]{McLerran:2007qj}%
  \BibitemOpen
  \bibfield  {author} {\bibinfo {author} {\bibfnamefont {L.}~\bibnamefont
  {McLerran}}\ and\ \bibinfo {author} {\bibfnamefont {R.~D.}\ \bibnamefont
  {Pisarski}},\ }\href {\doibase 10.1016/j.nuclphysa.2007.08.013} {\bibfield
  {journal} {\bibinfo  {journal} {Nucl. Phys.}\ }\textbf {\bibinfo {volume}
  {A796}},\ \bibinfo {pages} {83} (\bibinfo {year} {2007})},\ \Eprint
  {http://arxiv.org/abs/0706.2191} {arXiv:0706.2191 [hep-ph]} \BibitemShut
  {NoStop}%
\bibitem [{\citenamefont {Fujimoto}\ \emph {et~al.}(2024)\citenamefont
  {Fujimoto}, \citenamefont {Kojo},\ and\ \citenamefont
  {McLerran}}]{Fujimoto:2023mzy}%
  \BibitemOpen
  \bibfield  {author} {\bibinfo {author} {\bibfnamefont {Y.}~\bibnamefont
  {Fujimoto}}, \bibinfo {author} {\bibfnamefont {T.}~\bibnamefont {Kojo}}, \
  and\ \bibinfo {author} {\bibfnamefont {L.~D.}\ \bibnamefont {McLerran}},\
  }\href {\doibase 10.1103/PhysRevLett.132.112701} {\bibfield  {journal}
  {\bibinfo  {journal} {Phys. Rev. Lett.}\ }\textbf {\bibinfo {volume} {132}},\
  \bibinfo {pages} {112701} (\bibinfo {year} {2024})},\ \Eprint
  {http://arxiv.org/abs/2306.04304} {arXiv:2306.04304 [nucl-th]} \BibitemShut
  {NoStop}%
\bibitem [{\citenamefont {McLerran}\ and\ \citenamefont
  {Reddy}(2019)}]{McLerran:2018hbz}%
  \BibitemOpen
  \bibfield  {author} {\bibinfo {author} {\bibfnamefont {L.}~\bibnamefont
  {McLerran}}\ and\ \bibinfo {author} {\bibfnamefont {S.}~\bibnamefont
  {Reddy}},\ }\href {\doibase 10.1103/PhysRevLett.122.122701} {\bibfield
  {journal} {\bibinfo  {journal} {Phys. Rev. Lett.}\ }\textbf {\bibinfo
  {volume} {122}},\ \bibinfo {pages} {122701} (\bibinfo {year} {2019})},\
  \Eprint {http://arxiv.org/abs/1811.12503} {arXiv:1811.12503 [nucl-th]}
  \BibitemShut {NoStop}%
\bibitem [{\citenamefont {Koch}\ \emph {et~al.}(2024)\citenamefont {Koch},
  \citenamefont {McLerran}, \citenamefont {Miller},\ and\ \citenamefont
  {Vovchenko}}]{Koch:2024zag}%
  \BibitemOpen
  \bibfield  {author} {\bibinfo {author} {\bibfnamefont {V.}~\bibnamefont
  {Koch}}, \bibinfo {author} {\bibfnamefont {L.}~\bibnamefont {McLerran}},
  \bibinfo {author} {\bibfnamefont {G.~A.}\ \bibnamefont {Miller}}, \ and\
  \bibinfo {author} {\bibfnamefont {V.}~\bibnamefont {Vovchenko}},\ }\href@noop
  {} {\  (\bibinfo {year} {2024})},\ \Eprint {http://arxiv.org/abs/2403.15375}
  {arXiv:2403.15375 [nucl-th]} \BibitemShut {NoStop}%
\bibitem [{\citenamefont {de~Teramond}\ \emph {et~al.}(2018)\citenamefont
  {de~Teramond}, \citenamefont {Liu}, \citenamefont {Sufian}, \citenamefont
  {Dosch}, \citenamefont {Brodsky},\ and\ \citenamefont
  {Deur}}]{deTeramond:2018ecg}%
  \BibitemOpen
  \bibfield  {author} {\bibinfo {author} {\bibfnamefont {G.~F.}\ \bibnamefont
  {de~Teramond}}, \bibinfo {author} {\bibfnamefont {T.}~\bibnamefont {Liu}},
  \bibinfo {author} {\bibfnamefont {R.~S.}\ \bibnamefont {Sufian}}, \bibinfo
  {author} {\bibfnamefont {H.~G.}\ \bibnamefont {Dosch}}, \bibinfo {author}
  {\bibfnamefont {S.~J.}\ \bibnamefont {Brodsky}}, \ and\ \bibinfo {author}
  {\bibfnamefont {A.}~\bibnamefont {Deur}} (\bibinfo {collaboration} {HLFHS}),\
  }\href {\doibase 10.1103/PhysRevLett.120.182001} {\bibfield  {journal}
  {\bibinfo  {journal} {Phys. Rev. Lett.}\ }\textbf {\bibinfo {volume} {120}},\
  \bibinfo {pages} {182001} (\bibinfo {year} {2018})},\ \Eprint
  {http://arxiv.org/abs/1801.09154} {arXiv:1801.09154 [hep-ph]} \BibitemShut
  {NoStop}%
\bibitem [{\citenamefont {Sufian}\ \emph {et~al.}(2017)\citenamefont {Sufian},
  \citenamefont {de~T\'eramond}, \citenamefont {Brodsky}, \citenamefont
  {Deur},\ and\ \citenamefont {Dosch}}]{Sufian:2016hwn}%
  \BibitemOpen
  \bibfield  {author} {\bibinfo {author} {\bibfnamefont {R.~S.}\ \bibnamefont
  {Sufian}}, \bibinfo {author} {\bibfnamefont {G.~F.}\ \bibnamefont
  {de~T\'eramond}}, \bibinfo {author} {\bibfnamefont {S.~J.}\ \bibnamefont
  {Brodsky}}, \bibinfo {author} {\bibfnamefont {A.}~\bibnamefont {Deur}}, \
  and\ \bibinfo {author} {\bibfnamefont {H.~G.}\ \bibnamefont {Dosch}},\ }\href
  {\doibase 10.1103/PhysRevD.95.014011} {\bibfield  {journal} {\bibinfo
  {journal} {Phys. Rev. D}\ }\textbf {\bibinfo {volume} {95}},\ \bibinfo
  {pages} {014011} (\bibinfo {year} {2017})},\ \Eprint
  {http://arxiv.org/abs/1609.06688} {arXiv:1609.06688 [hep-ph]} \BibitemShut
  {NoStop}%
\bibitem [{\citenamefont {Lai}\ \emph {et~al.}(2010)\citenamefont {Lai},
  \citenamefont {Guzzi}, \citenamefont {Huston}, \citenamefont {Li},
  \citenamefont {Nadolsky}, \citenamefont {Pumplin},\ and\ \citenamefont
  {Yuan}}]{Lai:2010vv}%
  \BibitemOpen
  \bibfield  {author} {\bibinfo {author} {\bibfnamefont {H.-L.}\ \bibnamefont
  {Lai}}, \bibinfo {author} {\bibfnamefont {M.}~\bibnamefont {Guzzi}}, \bibinfo
  {author} {\bibfnamefont {J.}~\bibnamefont {Huston}}, \bibinfo {author}
  {\bibfnamefont {Z.}~\bibnamefont {Li}}, \bibinfo {author} {\bibfnamefont
  {P.~M.}\ \bibnamefont {Nadolsky}}, \bibinfo {author} {\bibfnamefont
  {J.}~\bibnamefont {Pumplin}}, \ and\ \bibinfo {author} {\bibfnamefont
  {C.~P.}\ \bibnamefont {Yuan}},\ }\href {\doibase 10.1103/PhysRevD.82.074024}
  {\bibfield  {journal} {\bibinfo  {journal} {Phys. Rev. D}\ }\textbf {\bibinfo
  {volume} {82}},\ \bibinfo {pages} {074024} (\bibinfo {year} {2010})},\
  \Eprint {http://arxiv.org/abs/1007.2241} {arXiv:1007.2241 [hep-ph]}
  \BibitemShut {NoStop}%
\bibitem [{\citenamefont {Brodsky}\ \emph {et~al.}(1998)\citenamefont
  {Brodsky}, \citenamefont {Pauli},\ and\ \citenamefont
  {Pinsky}}]{Brodsky:1997de}%
  \BibitemOpen
  \bibfield  {author} {\bibinfo {author} {\bibfnamefont {S.~J.}\ \bibnamefont
  {Brodsky}}, \bibinfo {author} {\bibfnamefont {H.-C.}\ \bibnamefont {Pauli}},
  \ and\ \bibinfo {author} {\bibfnamefont {S.~S.}\ \bibnamefont {Pinsky}},\
  }\href {\doibase 10.1016/S0370-1573(97)00089-6} {\bibfield  {journal}
  {\bibinfo  {journal} {Phys. Rept.}\ }\textbf {\bibinfo {volume} {301}},\
  \bibinfo {pages} {299} (\bibinfo {year} {1998})},\ \Eprint
  {http://arxiv.org/abs/hep-ph/9705477} {arXiv:hep-ph/9705477} \BibitemShut
  {NoStop}%
\bibitem [{\citenamefont {Angeles-Martinez}\ \emph {et~al.}(2015)\citenamefont
  {Angeles-Martinez} \emph {et~al.}}]{Angeles-Martinez:2015sea}%
  \BibitemOpen
  \bibfield  {author} {\bibinfo {author} {\bibfnamefont {R.}~\bibnamefont
  {Angeles-Martinez}} \emph {et~al.},\ }\href {\doibase
  10.5506/APhysPolB.46.2501} {\bibfield  {journal} {\bibinfo  {journal} {Acta
  Phys. Polon. B}\ }\textbf {\bibinfo {volume} {46}},\ \bibinfo {pages} {2501}
  (\bibinfo {year} {2015})},\ \Eprint {http://arxiv.org/abs/1507.05267}
  {arXiv:1507.05267 [hep-ph]} \BibitemShut {NoStop}%
\bibitem [{\citenamefont {Geesaman}\ \emph {et~al.}(1995)\citenamefont
  {Geesaman}, \citenamefont {Saito},\ and\ \citenamefont
  {Thomas}}]{Geesaman95}%
  \BibitemOpen
  \bibfield  {author} {\bibinfo {author} {\bibfnamefont {D.}~\bibnamefont
  {Geesaman}}, \bibinfo {author} {\bibfnamefont {K.}~\bibnamefont {Saito}}, \
  and\ \bibinfo {author} {\bibfnamefont {A.}~\bibnamefont {Thomas}},\
  }\href@noop {} {\bibfield  {journal} {\bibinfo  {journal} {Ann. Rev. Nucl.
  and Part. Sci.}\ }\textbf {\bibinfo {volume} {45}},\ \bibinfo {pages} {337}
  (\bibinfo {year} {1995})}\BibitemShut {NoStop}%
\bibitem [{\citenamefont {Norton}(2003)}]{Norton:2003cb}%
  \BibitemOpen
  \bibfield  {author} {\bibinfo {author} {\bibfnamefont {P.~R.}\ \bibnamefont
  {Norton}},\ }\href {\doibase 10.1088/0034-4885/66/8/201} {\bibfield
  {journal} {\bibinfo  {journal} {Rept. Prog. Phys.}\ }\textbf {\bibinfo
  {volume} {66}},\ \bibinfo {pages} {1253} (\bibinfo {year}
  {2003})}\BibitemShut {NoStop}%
\bibitem [{\citenamefont {Hen}\ \emph {et~al.}(2013)\citenamefont {Hen},
  \citenamefont {Higinbotham}, \citenamefont {Miller}, \citenamefont
  {Piasetzky},\ and\ \citenamefont {Weinstein}}]{Hen:2013oha}%
  \BibitemOpen
  \bibfield  {author} {\bibinfo {author} {\bibfnamefont {O.}~\bibnamefont
  {Hen}}, \bibinfo {author} {\bibfnamefont {D.~W.}\ \bibnamefont
  {Higinbotham}}, \bibinfo {author} {\bibfnamefont {G.~A.}\ \bibnamefont
  {Miller}}, \bibinfo {author} {\bibfnamefont {E.}~\bibnamefont {Piasetzky}}, \
  and\ \bibinfo {author} {\bibfnamefont {L.~B.}\ \bibnamefont {Weinstein}},\
  }\href {\doibase 10.1142/S0218301313300178} {\bibfield  {journal} {\bibinfo
  {journal} {Int. J. Mod. Phys. E}\ }\textbf {\bibinfo {volume} {22}},\
  \bibinfo {pages} {1330017} (\bibinfo {year} {2013})},\ \Eprint
  {http://arxiv.org/abs/1304.2813} {arXiv:1304.2813 [nucl-th]} \BibitemShut
  {NoStop}%
\bibitem [{\citenamefont {Hen}\ \emph {et~al.}(2017)\citenamefont {Hen},
  \citenamefont {Miller}, \citenamefont {Piasetzky},\ and\ \citenamefont
  {Weinstein}}]{Hen:2016kwk}%
  \BibitemOpen
  \bibfield  {author} {\bibinfo {author} {\bibfnamefont {O.}~\bibnamefont
  {Hen}}, \bibinfo {author} {\bibfnamefont {G.~A.}\ \bibnamefont {Miller}},
  \bibinfo {author} {\bibfnamefont {E.}~\bibnamefont {Piasetzky}}, \ and\
  \bibinfo {author} {\bibfnamefont {L.~B.}\ \bibnamefont {Weinstein}},\ }\href
  {\doibase 10.1103/RevModPhys.89.045002} {\bibfield  {journal} {\bibinfo
  {journal} {Rev. Mod. Phys.}\ }\textbf {\bibinfo {volume} {89}},\ \bibinfo
  {pages} {045002} (\bibinfo {year} {2017})},\ \Eprint
  {http://arxiv.org/abs/1611.09748} {arXiv:1611.09748 [nucl-ex]} \BibitemShut
  {NoStop}%
\bibitem [{\citenamefont {Miller}\ and\ \citenamefont
  {Smith}(2002)}]{Miller:2001tg}%
  \BibitemOpen
  \bibfield  {author} {\bibinfo {author} {\bibfnamefont {G.~A.}\ \bibnamefont
  {Miller}}\ and\ \bibinfo {author} {\bibfnamefont {J.~R.}\ \bibnamefont
  {Smith}},\ }\href {\doibase 10.1103/PhysRevC.66.049903} {\bibfield  {journal}
  {\bibinfo  {journal} {Phys. Rev. C}\ }\textbf {\bibinfo {volume} {65}},\
  \bibinfo {pages} {015211} (\bibinfo {year} {2002})},\ \bibinfo {note}
  {[Erratum: Phys.Rev.C 66, 049903 (2002)]},\ \Eprint
  {http://arxiv.org/abs/nucl-th/0107026} {arXiv:nucl-th/0107026} \BibitemShut
  {NoStop}%
\bibitem [{\citenamefont {Bohr}\ and\ \citenamefont
  {Motttelson}(1998)}]{Bohr1969}%
  \BibitemOpen
  \bibfield  {author} {\bibinfo {author} {\bibfnamefont {A.}~\bibnamefont
  {Bohr}}\ and\ \bibinfo {author} {\bibfnamefont {B.~R.}\ \bibnamefont
  {Motttelson}},\ }\href@noop {} {\emph {\bibinfo {title} {Nuclear Sturcture,
  Volume 1}}}\ (\bibinfo  {publisher} {World Scientific Publishing},\ \bibinfo
  {address} {Singapore},\ \bibinfo {year} {1998})\BibitemShut {NoStop}%
\end{thebibliography}
\end{document}